\documentstyle[11pt]{article}

\begin{document}

\title{
{\bf Comments on ``Another view on the velocity at the Schwarzschild horizon''}}

\author{ Abhas Mitra \\
{\em Theoretical Physics Division, BARC} \\ {\em Mumbai -400085,
India, amitra@apsara.barc.ernet.in} }
\date{}
\maketitle

\begin{abstract}
It is shown that  the conclusions reached by Tereno are completely faulty

\end{abstract}

We have recently shown that the Kruskal derivative assumes a form
\cite{mitra1}
\begin{equation}
{du\over dv} \rightarrow {f(r,t, dr/dt)\over \pm f(r,t, dr/dt)}
\end{equation}

because $u \rightarrow \pm v$ as $r \rightarrow 2M$. Although this limit
attains a value of $\pm 1$ irrespective of $f\rightarrow 0, \infty, or ~
anything$, Tereno \cite{tereno} refuses to accept this. Although, we have already
pointed out  that one should work out the limiting values of the
relevant fractions appropriately\cite{mitra2}, Tereno has decided to adopt
another view point on this issue\cite{tereno2}.

In his new note\cite{tereno2}, he has correctly reexpressed our result in
terms of the physical speed $V$, as seen by the Kruskal observer, and more
explicit Sch. relationships:

 For $r>2m$, the expression is,

\begin{equation}
\label{V}
V={{1+\tanh(t/4M){{dt}\over{dr}}(1-2M/r)}
\over{\tanh(t/4M)+{{dt}\over{dr}}(1-2M/r)}},
\end{equation}

Now since as  $r\rightarrow 2M$, $t\rightarrow \infty$ and $\tanh
(t/4M) \rightarrow 1$, the above equation approaches a form:

\begin{equation}
V = {f(r,t, dt/dr)\over f(r,t, dt/dr)}; \qquad r\rightarrow 2M
\end{equation}

Clearly, the foregoing limit assumes a value of 1 irrespective of whether
$f \rightarrow 0, \infty, or ~ anything$, Tereno thinks it is less than unity!
He on the other hand invokes (correctly) the expression for $dt/dr$ for a
radial geodesic:

\begin{equation}
\label{dtdr}
{{dt} \over {dr}}=-E\left(1-{{2M} \over {r}}\right)^{-1}
\left[ E^2-\left(1-{{2M} \over {r}}\right) \right]^{-1/2}.
\end{equation}

where $E$ is the conserved energy per unit rest mass.
It follows from this equation that

\begin{equation}
(1- 2M/r) {dt\over dr} = -E \left[ E^2-\left(1-{{2M} \over {r}}\right) \right]^{-1/2}
\end{equation}

Therefore, as $r\rightarrow 2M$, we have

\begin{equation}
(1- 2M/r) {dt\over dr} \rightarrow -1
\end{equation}

And if we put this result into Eq. (2), we will obtain

\begin{equation}
V={1-\tanh(t/4M)
\over -\tanh(t/4M)+ 1}; \qquad r\rightarrow 2M
\end{equation}

And clearly this above limit is again -1. But again, Tereno will not accept
i! Instead, he attempts to find an explicit $t=t(r)$ relation by a {\em
completely incorrect ansatz}. First he considers an approximate value of
the quantity in square bracket in Eq.(4). And when this approximation
 is valid in the
{\em infinetisimal neighbourhood of} $r=2M$, he, incorrectly integrates it
{\em over a finite region}. By feeding the resultant incorrect value of
$t(r)$ in Eq. (2) and by plotting the same he concludes that $V <1$. Even
if he is determined not to evaluate the appropriate limits and verify that
$v=1$ at $r=2M$, his later exercise was unnecessary because {\em the
precise and correct}
$t-r$ relationship is already known. For instance, he may look into Eq. (12.4.24)
of Shapiro \& Teukolsky\cite{shapiro}, we can write

\begin{equation}
{t\over 2M} = \ln\mid {x+1\over x-1}\mid + \left({R\over 2M} -1\right)
\left[ \eta + \left({R\over 4M}\right) (\eta +\sin \eta)\right]
\end{equation}

where $R$ is the value of $r$ at $t=0$ and
the ``cyclic coordinate'' $\eta$ is defined by

\begin{equation}
r= {R\over 2} (1 +\cos \eta)
\end{equation}

and the auxiliary variable

\begin{equation}
x= \left({R/2M -1\over R/r -1}\right)
\end{equation}

Now in principle using this exact parametric form of $t(r)$ and using the
exact form of $dt/dr$, one can plot Eq. (2). And then subject to the
numerical precision (note $t =\infty$ at $r=2M$), one may indeed verify that
$V=1$ at $r=2M$. However, since, $\tanh(t/4M) =1$ at $r=2M$, essentially,
we would be back to our starting position Eq. (1) by this procedure.

\vskip 1cm
Now let us also consider the ``Janis coordinates'' considered by Tereno.
Here the radial coordinate is

\begin{equation}
x_1 = (w+r)/\sqrt{2}
\end{equation}

and the time coordinate is

\begin{equation}
x_0 = (w-r)/\sqrt{2}
\end{equation}
where

\begin{equation}
w(r,t) = t +r + 2M \ln \mid {r-2M\over 2M}\mid
\end{equation}

As correctly indicated by Tereno, the physical speed measured in this
coordinate is $V_j = dx_1/dx_0$. And, in a general manner this can be
written as

\begin{equation}
V_j = {dx_1\over dx_0} = {dw/dt + dr/dt\over dw/dt -dr/dt}
\end{equation}

But if we go back to Eq. (4), it is found that

\begin{equation}
{dr\over dt} =0; \qquad r=2M
\end{equation}

Therefore, as $r\rightarrow 2M$, we have

\begin{equation}
V_j \rightarrow {dw/dt \over dw/dt} =1; \qquad r\rightarrow 2M
\end{equation}

And the eventual expression obtained in Eq. (13-14) of
Tereno\cite{tereno2} is simply incorrect. 

\vskip 1cm
If the reader is not still convinced about our result, we would remind a
basic relationship obtained by the Kruskal coordinates:

\begin{equation}
u^2 -v^2 = (r/2M -1) e^{r/2M}
\end{equation}

By differentiating both sides of this equation w.r.t., we obtain
\begin{equation}
2u {du\over dt} - 2v {dv\over dt} = \left[ {(r/2M -1)\over 2M} e^{r/2M} +
{e^{r/2M}\over 2M}\right] {dr\over dt}
\end{equation}

From  Eq. (4) , we note that $dr/dt=0$ at the EH, and therefore, the
foregoing equation yields
 \begin{equation}
 {du\over dt} = {v\over u}; \qquad r=2M
 \end{equation}
 
 But from Eq. (17), we find that $v/u =\pm 1$ at $r=2M$, and therefore
 \begin{equation}
 {du\over dt} = {v\over u} =\pm 1; \qquad r=2M
 \end{equation}

We have already explained why the value of $V\equiv 1$ at $r=2M$ in any
coordinate system. If the free fall speed measured by a Sch. observer is
$V_S$ and the relative velocity of the ``other static observer'' is
$V_{S-O}$ with respect to the Sch. observer, then we will have (locally):

\begin{equation}
V = {V_S \pm V_{S-O}\over 1 \pm V_S V_{S-O}}
\end{equation}

And since, $V_S =1$ at $r=2M$, we will have $\mid V\mid \equiv 1$.  We
hope Tereno will now realize that, indeed, $V=1$ at the event horizon.
And correspondingly, the geodesic of a material particle becomes null at
the EH. This in turn, implies that, there can not any finite mass BH, and
the collapse process continues indefinitely. For an overall scenario
see\cite{mitra3,mitra4} In case Tereno flashes another manuscript on the
same line, we shall not respond any further.


\bigskip \bigskip

\begin{thebibliography}{99}

\bibitem{mitra1}
Mitra, A. (1999), astro-ph/9904162

\bibitem{tereno}
Tereno,I. (1999), astro-ph/9905144

\bibitem{mitra2}
Mitra, A. (1999), astro-ph/9905175

\bibitem{tereno2}
Tereno, I (1999), astro-ph/9905298

\bibitem{shapiro}
Shapiro, S.L. and Teukolsky, S.A., ``Black Holes, White Dwarfs, and
Neutron Stars'' (Wiley, New York, 1983)


\bibitem{mitra3} Mitra, A. (1998), gr-qc/9810038

\bibitem{mitra4} Mitra, A. (1999), hep-th/9905182
\end{thebibliography}
\end{document}